\documentclass[pra, aps, twocolumn, superscriptaddress, showpacs, amsmath, amssymb]{revtex4}

\usepackage{graphicx}
\usepackage{float}
\usepackage{verbatim}
\usepackage[usenames,dvipsnames]{color}

\begin{document}

% MACROS
%%%%%%%%%%%%%%%%%%%%%%%%%%%%%%%%%%%%%%%%%%%%%%%%%%%%%%%%%%%%%%%%%%%%%

\newcommand{\beq}{\begin{equation}}
\newcommand{\eeq}{\end{equation}}
\newcommand{\beqa}{\begin{eqnarray}}
\newcommand{\eeqa}{\end{eqnarray}}
\newcommand{\lf}{\hfil \break \break}
\newcommand{\ahat}{\hat{a}}
\newcommand{\adag}{\hat{a}^{\dagger}}
\newcommand{\adagg}{\hat{a}_g^{\dagger}}
\newcommand{\bhat}{\hat{b}}
\newcommand{\bdag}{\hat{b}^{\dagger}}
\newcommand{\bdagg}{\hat{b}_g^{\dagger}}
\newcommand{\chat}{\hat{c}}
\newcommand{\cdag}{\hat{c}^{\dagger}}
\newcommand{\Pihat}{\hat{\Pi}}
\newcommand{\rhohat}{\hat{\rho}}
\newcommand{\shat}{\hat{\sigma}}
\newcommand{\ket}[1]{\mbox{$|#1\rangle$}}
\newcommand{\bra}[1]{\mbox{$\langle#1|$}}
\newcommand{\ketbra}[2]{\mbox{$|#1\rangle \langle#2|$}}
\newcommand{\braket}[2]{\mbox{$\langle#1|#2\rangle$}}
\newcommand{\bracket}[3]{\mbox{$\langle#1|#2|#3\rangle$}}
\newcommand{\mat}[1]{\overline{\overline{#1}}}
\newcommand{\hak}[1]{\left[ #1 \right]}
\newcommand{\vin}[1]{\langle #1 \rangle}
\newcommand{\abs}[1]{\left| #1 \right|}
\newcommand{\tes}[1]{\left( #1 \right)}
\newcommand{\braces}[1]{\left\{ #1 \right\}}

% TITLE
%%%%%%%%%%%%%%%%%%%%%%%%%%%%%%%%%%%%%%%%%%%%%%%%%%%%%%%%%%%%%%%%%%%%%

\title{Deterministic teleportation using single-photon entanglement as a resource}

\author{Gunnar Bj\"{o}rk}
\affiliation{Department of Applied Physics, Royal Institute of Technology (KTH)\\
AlbaNova University Center, SE-106 91 Stockholm, Sweden}
\affiliation{Temporary affiliation: NORDITA, Roslagstullsbacken 23, 106 91 Stockholm, Sweden}
\author{Amine Laghaout}
\author{Ulrik L. Andersen}
\affiliation{Department of Physics, Technical University of Denmark, Building 309, 2800 Lyngby, Denmark}
\affiliation{Temporary affiliation: NORDITA, Roslagstullsbacken 23, 106 91 Stockholm, Sweden}

\date{\today}

\begin{abstract}
We outline a proof that teleportation with a single particle is in principle just as reliable as with two particles. We thereby hope to dispel the skepticism surrounding single-photon entanglement as a valid resource in quantum information. A deterministic Bell state analyzer is proposed which uses only classical resources, namely coherent states, a Kerr non-linearity, and a two-level atom.
\end{abstract}


\maketitle

% Introduction
%%%%%%%%%%%%%%%%%%%%%%%%%%%%%%%%%%%%%%%%%%%%%%%%%%%%%%%%%%%%%%%%%%%%%

\section{Introduction}

Ever since Tan, Walls and Collett \cite{Tan} articulated the notion of single-particle nonlocality \cite{Hardy, Peres}, controversy has surrounded the ability of a single particle to exhibit entanglement \cite{Greenberger, Vaidman, Pawlowski, Drezet, vanEnk06}. Proposals \cite{Moussa,Steinberg,Lee,Bjork,Lee2,Nha,Dunningham,Anisimov,Cooper,Lougovski,Heaney,Heaney4} and experiments \cite{Hessmo,Papp,Salart} demonstrating single photon entanglement, non-locality, and entanglement purification have been performed, but still, the prospect of teleportation using single photon entanglement as the quantum resource has not been  regarded on equal footing with teleportation schemes involving ``carrier'' particles in each of the entangled modes. Although single-photon teleportation has been discussed quite extensively, and demonstrated experimentally \cite{Lombardi}, its success rate has been limited to at most 50\%. This could be taken by the detractors of single-particle entanglement as an indication that the involvement of the vacuum state as an agent of correlations bears with it a fundamental limitation. We argue that this lack of determinism, however, is not due to an intrinsic shortcoming of single-particle teleportation, but rather to the difficulty of implementing a deterministic analyzer for the following Bell states:
\beqa
\ket{\psi^{\pm}} = \frac{1}{\sqrt{2}} \tes{\ket{01} \pm \ket{10}} \label{Bell0110} \\
\ket{\varphi^{\pm}} = \frac{1}{\sqrt{2}} \tes{\ket{00} \pm \ket{11}} \label{Bell0011}
\eeqa
where, e.g., $\ket{01} \equiv \ket{0}_{\mbox{\scriptsize A}}\otimes\ket{1}_{\mbox{\scriptsize B}}$ is the shorthand notation for vacuum in Alice's mode and a single photon in Bob's.

If one is restricted to linear optics, it has been shown that this difficulty is fundamental \cite{Lutkenhaus}, and that the success rate appeared to be limited to the above mentioned 50\% . Recently, Pavi\v{c}i\'{c} demonstrated that this thereshold can be raised asymptotically to 100\% if one uses conditional dynamics on the polarization degree of freedom of a two-photon Bell state \cite{Pavicic}. However, his scheme does not lend itself to single photon Bell states because the delocalization behavior of polarization at beam splitters is different  for the vacuum state than for a single photon state. E.g., the splitting of, say, a vertically polarized photon \ket{V} on a beam splitter leads to a nonlocal superposition $\frac{1}{\sqrt{2}}\tes{\ket{V,0}+\ket{0,V}}$ whereas the vacuum remains separable $\ket{0}\otimes\ket{0}$, suggesting once again the alleged shortcoming of the vacuum.

However, what we wish to demonstrate in this paper is that fundamentally, Nature makes no difference as to whether the entanglement needed to perform  teleportation is carried by one, or more than one, particle. Hence, we will allow any classical resource, linear or non-linear, but no additional quantum resources. The reader should be warned that although the scheme we outline below is certainly experimentally implementable, it will not be the most practical scheme to teleport a state. Our aim is simply to argue that, at the fundamental level, any task that can be done by a multi-particle entangled state can also be achieved by the isomorphic state with the vacuum state and single-particle state as the basis. Specifically, we show that deterministic teleportation can be achieved with such a state as the only quantum resource.

In the context of single particle entanglement, non-locality, and teleportation there has been a debate as to whether particle superselection rules preclude such effects for single particles \cite{Wechsler,Cunha,Heaney2,Heaney3}. After all, in order to detect some event one needs a detector ``click'', and such a ``click'' is inevitably associated with a particle. We shall see that superselection rules can be circumvented by the use of auxiliary, but classical systems with an indeterminate particle number. Essentially the same technique was  suggested in \cite{Heaney3}, using a Bose-Einstein condensate as the auxiliary system. Below we shall instead use coherent states, which by contrast are classical resources. These states have the property that even if a particle is removed from a highly excited coherent state, the state remains essentially the same. That is, the state, and the same state with a particle removed, are essentially (and to an arbitrary degree) indistinguishable. This will allow us to let the auxiliary system ``lend'' a particle to the entangled system, and hence ``hide'' the particle number information that otherwise may ruin the intended task.

In this article, we aim to assert the soundness of single-particle teleportation by outlining an experiment which, in principle, can identify any of the four Bell states deterministically and with arbitrary accuracy. The main challenge to this end is that the Bell states $\ket{\varphi^{\pm}}$ are not energy (or more generally, particle number) eigenstates. In particular, no linear-optical scheme can deterministically resolve their phases \cite{Calsamiglia}. We propose a way around this by storing the photonic qubits in two-level atoms. Owing to the two-dimensional Hilbert space of a two-level system, its stored qubits can conveniently be rotated on the Bloch sphere via coherent excitations. Once aligned with the energy eigenbasis of the atom, the orientation of the initial qubit can easily be deduced due to the unitarity of the rotation. Before treating the two-mode case of $\ket{\varphi^{\pm}}$, we first consider in Sec. \ref{sec:HadamardRotation} the Hadamard rotation of the single-mode qubit $\frac{1}{\sqrt{2}}\tes{\ket{0}\pm{1}} \rightarrow \braces{\ket{0}, \ket{1}}$. We then follow up in Sec. \ref{sec:Analyzer} with a description of the actual teleportation setup and its two-stage Bell analyzer.

% Phase resolution between vacuum and single photon
%%%%%%%%%%%%%%%%%%%%%%%%%%%%%%%%%%%%%%%%%%%%%%%%%%%%%%%%%%%%%%%%%%%%%

\section{Hadamard rotation of a vacuum-photon superposition}
\label{sec:HadamardRotation}

Consider a qubit made up of an equal superposition of the vacuum and a single photon:
\beq
\ket{X_{\theta}} = \frac{1}{\sqrt{2}} \tes{\ket{0} + e^{i\theta}\ket{1}} \label{eq:qubit}
\eeq
where $\theta$ is the equatorial angle on the Bloch sphere. Our first goal is to devise a projector $\Pihat_{\theta} = \ketbra{X_{\theta}}{X_{\theta}}$ which can resolve the phase $\theta$. Note that any projector $\Pihat_{\theta}$ can be implemented from any other $\Pihat_{\beta}$ by interposing a phase shift $\Delta\theta = \theta - \beta$. We will show in \ref{sec:PhiDiscrimination} that such a projector, when applied in parallel to the two modes of $\ket{\varphi^{\pm}}$, will allow us to resolve the sign of the superposition. We shall for now restrict ourselves to the single-mode case and describe how $\ket{X_{0}}$ can be distinguished from $\ket{X_{\pi}}$.

Let's define two initially separated Hilbert spaces pertaining to an atomic and a photonic mode, respectively. Formally, the space under consideration is $\mathcal{H} = \mathcal{H}_{\mbox{\scriptsize atom}}\otimes\mathcal{H}_{\mbox{\scriptsize photon}}$ where $\mathcal{H}_{\mbox{\scriptsize photon}} = \braces{\ket{n} : n \in \mathbb{N}}$ and $\mathcal{H}_{\mbox{\scriptsize photon}} = \braces{\ket{g}, \ket{e}}$. Here, $n$ denotes the number of photons, and $\ket{g}$ and $\ket{e}$ denote the ground and excited atomic states, respectively. The interaction between the two modes is dictated by the Jaynes-Cummings (JC) Hamiltonian, expressed below in the rotating wave approximation:
\beq
\hat{H} = \hbar\gamma \tes{\shat^{-}\adag + \shat^{+}\ahat} \label{eq:JCHamiltonian}
\eeq
where $\ahat$ ($\adag$) is the photon annihilation (creation) operator, and $\shat^{+}$ ($\shat^{-}$) is the atomic raising (lowering) operator. $\gamma$ quantifies the strength of the photon-atom coupling. The transformations undergone by any preparation in $\mathcal{H}$ under the action of $\hat{H}$ are summarized in the appendix and shall be used in what follows.

The candidate qubits $\ket{X_{0}}$ and $\ket{X_{\pi}}$ to be measured are initially stored in the photonic mode whereas the atom is prepared in the ground state. Upon an interaction time $\tau = \frac{\pi}{2\gamma}$, we obtain the transformation
\beq
\frac{1}{\sqrt{2}} \ket{g}\otimes\tes{\ket{0} \pm \ket{1}} \stackrel{\tau}{\longrightarrow} \frac{1}{\sqrt{2}} \tes{\ket{g} \mp i\ket{e}}\otimes\ket{0} \label{eq:photon2atom}
\eeq
whereby the photonic qubit has been transferred to the atomic mode and the state is are once again separable. The atomic qubit at this stage is not yet measurable in the energy eigenbasis. It can, however, be rotated so as to align itself with the eigenstates of the atom by shining a strong coherent beam $\ket{\alpha}$ with $\abs{\alpha} \gg 1$. (This coherent state will incidentally serve as a reference phase.) If one chooses an interaction time $t_{\mbox{\scriptsize s}} = \frac{\pi}{4\gamma\abs{\alpha}}$ the states transforms to a very good approximation (see below) as
\beq
\left\{
	\begin{array}{lr}
		\frac{1}{\sqrt{2}}\tes{\ket{g}+i\ket{e}}\otimes\ket{\alpha} \stackrel{t_{\mbox{\scriptsize s}}}{\longrightarrow} \ket{g}\otimes\ket{\alpha} \\
		\frac{1}{\sqrt{2}}\tes{\ket{g}-i\ket{e}}\otimes\ket{\alpha} \stackrel{t_{\mbox{\scriptsize s}}}{\longrightarrow} \ket{e}\otimes\ket{\alpha} \label{eq:Hadamard}
	\end{array}.
\right.
\eeq

If one now determines via, say, a fluorescence measurement that the final state of the atom was the ground (excited) state then one can conclude that the initial qubit was $\ket{X_{\pi}}$ ($\ket{X_{0}}$). A sketch of the physics underlying the transformations (\ref{eq:photon2atom}) and (\ref{eq:Hadamard}) is shown in Fig. \ref{fig:Projector}.

\begin{figure}
  \includegraphics[width=0.9\columnwidth]{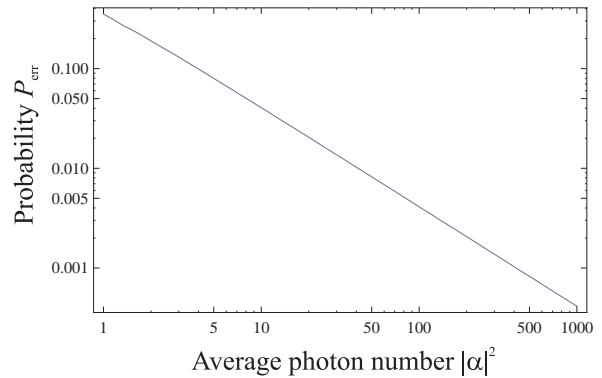}
  \caption{The probability $P_{\mbox{\scriptsize err}}$ of an erroneous projection as a function of the coherent state average photon number.}
\label{Fig:1}
\end{figure}

As derived in the appendix, however, an error in the correspondence between initial and final states in (\ref{eq:Hadamard}) will arise for weaker coherent fields. This is where the superselection rule kicks in, because it is clear that the left- and the right hand side of (\ref{eq:Hadamard}) do not contain the same number of particles on average. An exact analysis of the transformation, made in the Appendix, shows that the probability for such an error decreases with the strength $\abs{\alpha}$ of the coherent state. For example, the probability of  erroneously identifying $\ket{X_0}$ instead of $\ket{X_{\pi}}$ is given by
\beq
P_{\mbox{\scriptsize err}} = \frac{e^{-\abs{\alpha}^2}}{2} \sum_{n=0}^\infty \frac{\abs{\alpha}^{2n}}{n!} \abs{ \cos\tes{\frac{\pi\sqrt{n}}{4|\alpha|}} - \frac{\sqrt{n}}{|\alpha|} \sin\tes{\frac{\pi\sqrt{n}}{4|\alpha|}}}^2.
\eeq
 The fidelity of the Hadamard rotation is therefore contingent on the strength of the coherent $\frac{\pi}{2}$-pulses. The error probability is plotted in Fig. \ref{Fig:1}, and it can be seen that already for $|\alpha|^2 = 50$, the error probability is at the 1\% level. This means that already for rather modest coherent state excitations, the unitarity of the Hadamard operation in (\ref{eq:Hadamard}) is effectively achieved.

\begin{figure}[ht]
\includegraphics[scale=.65]{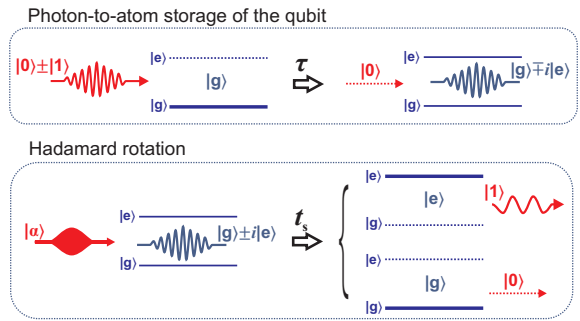}
\caption{(Color online) Sketch of the two-stage implementation of the $\Pihat_{\theta}$ projector for $\theta \in \braces{0, \pi}$. The photonic and atomic modes are colored in red and blue, respectively. First, the photonic qubit is transfered to an atom initially in the ground state (top). Second, a coherent $\frac{\pi}{2}$-pulse is applied on the atom so as to rotate the qubit into one of the energy basis vectors $\ket{g}$ or $\ket{e}$ (bottom). The subsequent de-excitation of the atom (or lack thereof) will reveal that the initial photonic qubit was $\ket{X_{0}}$ (or $\ket{X_{\pi}}$).}
\label{fig:Projector}
\end{figure}

% The Bell-state analyzer
%%%%%%%%%%%%%%%%%%%%%%%%%%%%%%%%%%%%%%%%%%%%%%%%%%%%%%%%%%%%%%%%%%%%%

\section{The Bell-state analyzer}
\label{sec:Analyzer}

\begin{figure}[h]
\includegraphics[scale=.8]{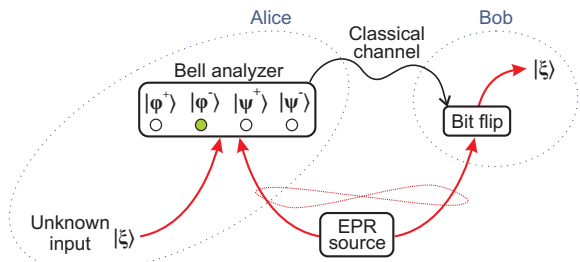}
\caption{(Color online) Teleportation protocol. (Here, the instance where Alice detects $\ket{\phi^{-}}$ is highlighted.)}
\label{fig:Teleportation}
\end{figure}

The teleportation protocol is sketched in Fig. \ref{fig:Teleportation}. It consists of an entangled resource  $\ket{\psi^{+}} = \frac{1}{\sqrt{2}}\tes{\ket{01}+\ket{10}}$ linking Alice and Bob and an unknown state $\ket{\xi} = a\ket{0} + b\ket{1}$ (where $\abs{a}^2+\abs{b}^2 = 1$) to be teleported from Alice to Bob. The overall tripartite state, with the first two modes belonging to Alice and the last to Bob, reads
\beqa
\ket{\Psi} & = & \ket{\xi}\otimes\ket{\psi^{+}} \nonumber\\
& = &  \frac{1}{\sqrt{2}}\tes{a\ket{001}+a\ket{010}+b\ket{101}+b\ket{110}} \nonumber\\
& = & \frac{1}{2}\ket{\varphi^{+}}\otimes\tes{a\ket{1}+b\ket{0}} + \frac{1}{2}\ket{\varphi^{-}}\otimes\tes{a\ket{1}-b\ket{0}} +\nonumber\\
& & \frac{1}{2}\ket{\psi^{+}}\otimes\tes{a\ket{0}+b\ket{1}} + \frac{1}{2}\ket{\psi^{-}}\otimes\tes{a\ket{0}-b\ket{1}}. \nonumber
\eeqa

Upon the detection of $\ket{\psi^{\pm}}$ or $\ket{\varphi^{\pm}}$, Alice can inform Bob via a classical channel that he has in his possession $a\ket{0}\pm b\ket{1}$ or $a\ket{1}\pm b\ket{0}$, respectively. Bob can then perform a local unitary operation of his qubit to recover $\ket{\xi}$ (Table \ref{tab:BobLocalOperation}).

\begin{table}
\caption{Table of the local operations to be performed by Bob based on the four possible Bell states measured by Alice. In the case where Alice measures $\ket{\psi^+}$, no action need to be taken by Bob. In the other three cases, he will have to apply a photonic $\pi$ phase shift (abridged `shift' below) and/or transfer the photonic qubit to a two-level atom and then apply a coherent $\pi$ pulse (abridged `transfer, flip').}
\begin{ruledtabular}
\begin{tabular}{ll}
Bell state & Local operation \\
\hline
$\ket{\varphi^+}$ & $a\ket{1} + b\ket{0} \rightarrow \mbox{transfer, flip} \rightarrow \ket{\xi}$ \\
$\ket{\varphi^-}$ & $a\ket{1} - b\ket{0} \rightarrow \mbox{shift, transfer, flip} \rightarrow \ket{\xi} $ \\
$\ket{\psi^+}$ & $a\ket{0} + b\ket{1} = \ket{\xi}$  \\
$\ket{\psi^-}$ & $a\ket{0} - b\ket{1} \rightarrow \mbox{shift} \rightarrow \ket{\xi}$  \\
\end{tabular}
\end{ruledtabular}
\label{tab:BobLocalOperation}
\end{table}

\begin{figure*}[ht]
\includegraphics[scale=.8]{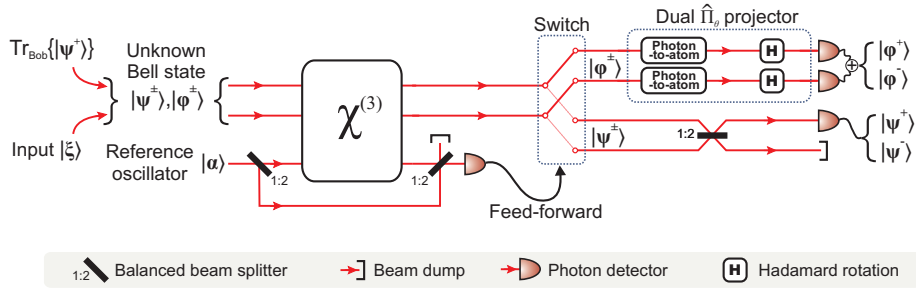}
\caption{(Color online) Conceptual sketch of the Bell analyzer. The first part consists of a quantum non-demolition measurement which separates $\ket{\psi^{\pm}}$ from $\ket{\varphi^{\pm}}$ by inducing a phase shift in an auxiliary coherent field if $\ket{\psi^{\pm}}$ was input. Depending on the outcome of this first measurement, a double pole, double throw (DPDT) switch forwards the state to either a balanced beam splitter (for $\ket{\psi^{\pm}}$), or a ``dual-rail'' extension of the projector discussed in Sec. \ref{sec:HadamardRotation} (for $\ket{\varphi^{\pm}}$).}
\label{fig:Analyzer}
\end{figure*}

Let's now focus on the Bell analyzer. We propose that it consist of two steps: Alice first distinguishes $\ket{\psi^{\pm}}$ from $\ket{\varphi^{\pm}}$, and then she determines the signs of each superposition with separate setups. We discuss each step in the two subsections below. A sketch of the Bell analyzer is shown in Fig. \ref{fig:Analyzer}.

% Discrimination between \psi and \phi
%%%%%%%%%%%%%%%%%%%%%%%%%%%%%%%%%%

\subsection{Discrimination between $\ket{\psi^{\pm}}$ and $\ket{\varphi^{\pm}}$}
\label{sec:PhiDiscrimination}

The main difference between $\ket{\psi^{\pm}}$ and $\ket{\varphi^{\pm}}$ is that the former are energy eigenstates which are easily separable by a rotation in the energy basis (e.g., with a 50/50 beam splitter). The latter, however, are not energy or particle eigenstates and thus require a more elaborate treatment to reveal the sign of their superposition by an energy (particle) counting detector. Our first task is therefore to branch off $\ket{\psi^{\pm}}$ and $\ket{\varphi^{\pm}}$ so that each be forwarded to the appropriate analyzer. We propose to achieve this sorting with a dual-rail quantum non-demolition (QND) measurement, first proposed for quantum error correction \cite{Chuang}. The idea behind this QND is to induce a phase shift in an auxiliary coherent beam depending on whether the total number of photons in the Bell state is odd (as in $\ket{\psi^{\pm}}$) or even (as in $\ket{\varphi^{\pm}}$). The coherent probe remains separable all along so that no collapse is incurred on the individual Bell states. The interaction Hamiltonian, which could be implemented physically as a cross-Kerr effect \cite{Imoto}, is written in the rotating wave approximation as
\beq
\hat{H}_{\mbox{\scriptsize int}} = \hbar\kappa\tes{\adag\ahat + \bdag\bhat}\cdag\chat
\eeq
where $\kappa$ is the strength of the interaction.

If we set the interaction time to $\tau_{\mbox{\scriptsize M}} = \frac{\pi}{\kappa}$, the propagator becomes $e^{-i\pi\tes{\adag\ahat + \bdag\bhat}\cdag\chat}$ and we are then faced with two possible scenarios. If the Bell state is $\ket{\psi^{\pm}}$, then the overall state $\ket{\psi^{\pm}}\otimes\ket{\alpha}$ transforms as follows:
\beqa
\ket{\psi^{\pm}}\otimes\ket{\alpha} & = & \frac{e^{\frac{\abs{\alpha}^2}{2}}}{\sqrt{2}}\tes{\ket{01}\pm\ket{10}}\otimes\sum_{n=0}^{\infty} \frac{\alpha^n}{\sqrt{n!}}\ket{n} \nonumber\\
& \stackrel{\tau_{\mbox{\scriptsize M}}}{\longrightarrow} & \frac{e^{\frac{\abs{\alpha}^2}{2}}}{\sqrt{2}}\tes{\ket{01}\pm\ket{10}}\otimes\sum_{n=0}^{\infty} e^{-i\pi n} \frac{\alpha^n}{\sqrt{n!}}\ket{n} \nonumber\\
& = & \frac{1}{\sqrt{2}}\tes{\ket{01}\pm\ket{10}}\otimes\ket{-\alpha} \nonumber\\
& = & \ket{\psi^{\pm}}\otimes\ket{-\alpha}.
\eeqa

On the other hand, if we start with $\ket{\varphi^{\pm}}$, a similar derivation leaves the state unchanged:
\beq
\ket{\varphi^{\pm}}\otimes\ket{\alpha} \stackrel{\tau_{\mbox{\scriptsize M}}}{\longrightarrow} \ket{\varphi^{\pm}}\otimes\ket{\alpha}.
\eeq

It can now be seen that the differentiation of $\ket{\psi^{\pm}}$ and $\ket{\varphi^{\pm}}$ can be achieved be comparing the phases of the auxiliary coherent fields: Only those coherent fields that have interacted with $\ket{\psi^{\pm}}$ acquire a $\pi$ phase shift, those that interacted with $\ket{\varphi^{\pm}}$ remain unchanged. The acquisition of the $\pi$ phase shift can be observed by a simple classical interference between the probe beam and a reference coherent state on a balanced beam splitter.

% Discrimination between \phi+ and \phi-
%%%%%%%%%%%%%%%%%%%%%%%%%%%%%%%%%%

\subsection{Discrimination between $\ket{\varphi^{+}}$ and \ket{\varphi^{-}}}

Now that $\ket{\psi^{\pm}}$ and $\ket{\varphi^{\pm}}$ are branched off, there remain to determine the signs at the superscript in either case. As already mentioned, the differentiation between the triplet ($\psi^{+}$) and singlet states ($\psi^{-}$) can be done easily by joining the two constituent modes on a balanced beam splitter. The outgoing modes become disentangled and the detection of the photon at either output ports has a direct correspondence to the sign of the superposition.

A more complicated situation occurs when the QND announces the states $\ket{\varphi^{\pm}}$. The discrimination between the signs requires a two-mode extension to the single-mode projector described in Sec. \ref{sec:HadamardRotation}. This is done by placing a ground-state atom in the path of each of the incoming photonic qubits. After an interaction time of $\tau = \frac{\pi}{2\gamma}$, the evolution of the photon-atom system will be the straightforward extension of (\ref{eq:photon2atom}), namely
\beqa
\ket{gg}\otimes\ket{\varphi^{\pm}} & = & \frac{1}{\sqrt{2}}\ket{gg}\otimes\tes{\ket{00}\pm\ket{11}} \nonumber\\
& \stackrel{\tau}{\longrightarrow} & \frac{1}{\sqrt{2}}\tes{\ket{gg}\mp\ket{ee}}\otimes\ket{00}.
\eeqa

Now that the qubits have been completely transferred from the photonic to the atomic modes, we can call upon the Hadamard transformation worked out in (\ref{eq:Hadamard}). This rotation gives:
\beq
\left\{
	\begin{array}{lr}
		\frac{1}{\sqrt{2}} \tes{\ket{gg}-\ket{ee}}\otimes\ket{\alpha\alpha} \stackrel{t_{\mbox{\scriptsize s}}}{\longrightarrow} \frac{1}{\sqrt{2}}\tes{\ket{gg}+\ket{ee}}\otimes\ket{\alpha\alpha} \\
		\frac{1}{\sqrt{2}} \tes{\ket{gg}+\ket{ee}}\otimes\ket{\alpha\alpha} \stackrel{t_{\mbox{\scriptsize s}}}{\longrightarrow} \frac{1}{\sqrt{2}}\tes{\ket{ge}+\ket{eg}}\otimes\ket{\alpha\alpha}
	\end{array}
\right.
\eeq
where once again $t_{\mbox{\scriptsize s}} = \frac{\pi}{4\gamma\abs{\alpha}}$ is the time it takes to apply a $\frac{\pi}{2}$-coherent pulse.

The difference between the two final states lies in the parity of the energy quanta stored in the atoms. An initial photonic state $\ket{\varphi^{-}}$ corresponds to a total energy of exactly one quantum: $\frac{1}{\sqrt{2}}\tes{\ket{ge}+\ket{eg}}$. Conversely, $\ket{\varphi^{+}}$ leads to either zero or two quanta: $\frac{1}{\sqrt{2}}\tes{\ket{gg}+\ket{ee}}$. The efficiency of this Hadamard rotation, as argued in Sec. \ref{sec:HadamardRotation}, increases with the mean photon number of the $\frac{\pi}{2}$ pulses and can thus be made asymptotically ideal for strong coherent fields.

% Deterministic quantum computing
%%%%%%%%%%%%%%%%%%%%%%%%%%%%%%%%%%%%%%%%%%%%%%%%%%%%%%%%%%%%%%%%%%%%%

\section{Deterministic quantum computing}
\label{sec:DeterministicQC}

We finally show that using the experimental techniques presented in this paper, it is also possible to achieve deterministic quantum computing based on single photon entanglement. It has been shown by Lund and Ralph \cite{Lund} that nondeterministic quantum computing using the superposition of vacuum and a single photon as a qubit can be obtained with linear optics and photon counters. However, by allowing for non-linear operations, it is possible to bring this idea into a deterministic setting.

A universal set of quantum gates could consist of the phase rotation gate, the Hadamard gate and the control sign shift (CS) gate. The phase rotation gate is easily implementable using a simple phase delay. A deterministic Hadamard gate can be constructed using the JC interaction as outlined in Sec. \ref{sec:HadamardRotation}. The CS gate can be implemented by storing the input modes $(a|0\rangle+b|1\rangle)\otimes(c|0\rangle+d|1\rangle)$ in a pair of atoms (via the JC interaction), applying a $\pi$ pulse and letting it decay:
\begin{eqnarray}
& & (ac|00\rangle+ ad|01\rangle+ bc|10\rangle+ bd|11\rangle)\otimes|gg\rangle \nonumber\\
& \stackrel{\tau}{\longrightarrow} & (ac|gg\rangle+ ad|ge\rangle+ bc|eg\rangle+ bd|ee\rangle)\otimes|00\rangle \nonumber\\
& \stackrel{2t_{\mbox{\scriptsize s}}}{\longrightarrow}& (ac|gg\rangle+ ad|ge\rangle+ bc|eg\rangle- bd|ee\rangle)\otimes|00\rangle \nonumber\\
& \stackrel{\mbox{\scriptsize decay}}{\longrightarrow} & (ac|00\rangle+ ad|01\rangle+ bc|10\rangle- bd|11\rangle)\otimes|gg\rangle \nonumber
\end{eqnarray}
Combining this CS gate with the above mentioned phase and Hadamard gate, universal quantum computation based on qubits of the form (\ref{eq:qubit}) can in principle be executed.

% Conclusion
%%%%%%%%%%%%%%%%%%%%%%%%%%%%%%%%%%%%%%%%%%%%%%%%%%%%%%%%%%%%%%%%%%%%%

\section{Conclusion}
\label{sec:Conclusion}

Our proposal is difficult to implement experimentally in that it requires expertise in two separate and highly specialized areas, namely the generation of single photons and the manipulation of light-matter interactions. We believe however that our theoretical sketch will help bring some closure to the debate that still surrounds the notion of single-particle nonlocality. The root of this debate can be traced to the perception of the vacuum $\ket{0}$ as a singular---if not pathological \cite{Pawlowski}---state whose similarity to the other Fock states has little physical meaning beyond mathematical isomorphism. By building on earlier discussions about the entanglement and non-locality of this state \cite{Hardy,Moussa,Cunha,Heaney3}, and by showing the full power of single-photon teleportation with no additional quantum resources, we hope to have proved the contrary.

\section*{Acknowledgments}
The authors acknowledge fruitful discussions with Prof. S. Kilin. The work was supported by the Swedish Research Council (VR) trough its Linn\ae us Center of Excellence ADOPT, and by the Danish Agency for Science, Technology and Innovation (FNU).

% Appendix
%%%%%%%%%%%%%%%%%%%%%%%%%%%%%%%%%%%%%%%%%%%%%%%%%%%%%%%%%%%%%%%%%%%%%

\appendix
\section{Jaynes-Cummings model}

We shall summarize here the mathematics behind the Hadamard rotation treated in Sec. \ref{sec:HadamardRotation}. This is based on an application of the Jaynes-Cummings model whose Hamiltonian has already been presented in (\ref{eq:JCHamiltonian}). The Schr\"{o}dinger equation corresponding to this system is solved by
\begin{widetext}
\beq
\ket{\psi(t)} = \sum_{n=0}^{\infty} \hak{\tes{c_{e}c_{n}\cos(\gamma t\sqrt{n+1}) - i c_{g}c_{n+1}\sin(\gamma t\sqrt{n+1})}\ket{e} + \tes{c_{g}c_{n}\cos(\gamma t\sqrt{n}) - i c_{e}c_{n-1}\sin(\gamma t\sqrt{n})}\ket{g}}\ket{n} \label{eq:psiSolution}
\eeq
\end{widetext}
where the initial state is given by
\beqa
\ket{\psi(0)} & = & \ket{\psi_{\mbox{\scriptsize atom}}(0)} \otimes \ket{\psi_{\mbox{\scriptsize photon}}(0)} \nonumber\\
& = & \tes{c_{g}\ket{g} + c_{e}\ket{e}} \otimes \sum_{n=0}^{\infty} c_{n} \ket{n},
\eeqa
and $c_{g}, c_{e}$, and $c_{n}$ are complex. An in-depth derviation of (\ref{eq:psiSolution}) is given in \cite{Gerry}.

Three key transformations of the atom-photon eigenstates are of interest to us, namely
\beqa
\ket{g,0} & \rightarrow & \ket{g,0}, \\
\ket{g,n} & \rightarrow & \cos(\gamma t\sqrt{n})\ket{g,n} - i\sin(\gamma t\sqrt{n})\ket{e,n-1}, \\
\ket{e,n} & \rightarrow & \cos(\gamma t\sqrt{n+1})\ket{e,n} - i\sin(\gamma t\sqrt{n+1})\ket{g,n+1}.
\eeqa

The transfer of the qubit from the photon to the atomic modes is thus given by
\beqa
& & \ket{g}\otimes\frac{1}{\sqrt{2}}\tes{\ket{0}\pm\ket{1}}  = \frac{1}{\sqrt{2}}\tes{\ket{g0}\pm\ket{g1}} \nonumber\\
& \rightarrow & \frac{1}{\sqrt{2}} \tes{\ket{g0} \mp i\sin(\gamma t)\ket{e0} \pm \cos(\gamma t)\ket{g1}} \nonumber\\
& = & \braces{t = \tau = \frac{\pi}{2\gamma}} \nonumber\\
& = & \frac{1}{\sqrt{2}}\tes{\ket{g}\mp i\ket{e}}\otimes\ket{0}
\eeqa

Now that the qubit is stored in the atomic mode, let's derive how a coherent excitation $\ket{\alpha}$ performs the Hadamard rotation.
\begin{widetext}
\beqa
& & \frac{1}{\sqrt{2}}\tes{\ket{g}\pm i\ket{e}}\otimes{\ket{\alpha}} = \frac{1}{\sqrt{2}}\tes{\ket{g}\pm i\ket{e}}\otimes e^{-\frac{\abs{\alpha}^2}{2}} \sum_{n=0}^{\infty} \frac{\alpha^n}{\sqrt{n!}}\ket{n} = \frac{e^{-\frac{\abs{\alpha}^2}{2}}}{\sqrt{2}} \sum_{n=0}^{\infty} \frac{\alpha^n}{\sqrt{n!}} \tes{\ket{gn}\pm i\ket{en}} \nonumber\\
& \rightarrow & \frac{e^{-\frac{\abs{\alpha}^2}{2}}}{\sqrt{2}} \sum_{n=0}^{\infty} \frac{\alpha^n}{\sqrt{n!}} \Big[\cos(\gamma t\sqrt{n})\ket{gn} -i\sin(\gamma t\sqrt{n})\ket{e,n-1} \pm i\cos(\gamma t\sqrt{n+1})\ket{en} \pm \sin(\gamma t\sqrt{n+1})\ket{g,n+1} \Big] \nonumber\\
& = & \frac{e^{-\frac{\abs{\alpha}^2}{2}}}{\sqrt{2}} \sum_{n=0}^{\infty} \frac{\alpha^n}{\sqrt{n!}}\Big[ \tes{\cos(\gamma t\sqrt{n}) \pm \frac{\sqrt{n}}{\alpha}\sin(\gamma t\sqrt{n})}\ket{gn} +i\tes{\pm\cos(\gamma t\sqrt{n+1}) - \frac{\alpha}{\sqrt{n+1}}\sin(\gamma t\sqrt{n+1})}\ket{en} \Big] \nonumber\\
& = & \braces{t = t_{\mbox{\scriptsize s}} = \frac{\pi}{4\gamma\abs{\alpha}}} \nonumber\\
& = & \frac{e^{-\frac{\abs{\alpha}^2}{2}}}{\sqrt{2}} \sum_{n=0}^{\infty} \frac{\alpha^n}{\sqrt{n!}}\hak{ \tes{\cos\tes{\frac{\pi\sqrt{n}}{4\abs{\alpha}}} \pm \frac{\sqrt{n}}{\alpha}\sin\tes{\frac{\pi\sqrt{n}}{4\abs{\alpha}}}}\ket{gn} +i\tes{\pm\cos\tes{\frac{\pi\sqrt{n+1}}{4\abs{\alpha}}} - \frac{\alpha}{\sqrt{n+1}}\sin\tes{\frac{\pi\sqrt{n+1}}{4\abs{\alpha}}}}\ket{en}}
\eeqa
\end{widetext}
If we now assume that $\alpha \approx n^{1/2}$, the cosine and sine function become approximately equal, thereby finalizing the Hadamard transformations (\ref{eq:Hadamard}). One needs however to keep track of the error arising from the approximation. For example, the probability of erroneously obtaining a final state $\ket{gn}$ instead of $\ket{en}$ will be given by any non-zero remnant in the factor of $\ket{gn}$:
\beq
P_{\mbox{\scriptsize err}} = \frac{e^{-\abs{\alpha}^2}}{2} \sum_{n=0}^\infty \frac{\abs{\alpha}^{2n}}{n!} \abs{ \cos\tes{\frac{\pi\sqrt{n}}{4|\alpha|}} - \frac{\sqrt{n}}{|\alpha|} \sin\tes{\frac{\pi\sqrt{n}}{4|\alpha|}}}^2.
\eeq

% Bibliography
%%%%%%%%%%%%%%%%%%%%%%%%%%%%%%%%%%%%%%%%%%%%%%%%%%%%%%%%%%%%%%%%%%%%%

\end{document}